# Visualizing "Fermi arcs" in the Weyl semimetal TaAs

Rajib Batabyal[1,†], Noam Morali[1,†], Nurit Avraham[1,†], Yan Sun[2], Marcus Schmidt[2], Claudia Felser[2], Ady Stern[1], Binghai Yan[2], Haim Beidenkopf[1]

[1]Department of Condensed Matter Physics, Weizmann Institute of Science, Rehovot 7610001, Israel

[2]Max Planck Institute for Chemical Physics of Solids, D-01187, Dresden, Germany

**One of the hallmarks of Weyl semi-metals is the existence of unusual topological surface states known as "Fermi arcs" [1-3]. The formation of these states is guaranteed by the existence of bulk Weyl points with opposite chirality. Tantalum Arsenide (TaAs) [4-9], a member of the newly discovered family of Weyl semi-metals [4,5], harbors a host of non-topological ("trivial") surface states overlapping in energy with the predicted 12 "Fermi arcs". This overlap poses a major challenge in identifying the signatures of the arcs [10]. Here we harness the inherently distinct spatial structure of trivial and Fermi arc states to visualize the Fermi arcs for the first time using scanning tunneling microscopy. We do so in four distinct ways, each of which highlights a different aspect of their unusual nature. We reveal their relatively isotropic scattering signature, their energy dispersion and its relation to the bulk Weyl points, their deep bulk penetration relative to that of non-topological surface states and their weak coupling to the atomic structure. The latter is obtained by accounting for the spatial structure of the Bloch wavefunction and its effect on the scattering properties of the electrons off lattice defects in general. It thus provides a novel analysis tool for the spectroscopic characterization of electronic wavefunctions using scanning tunneling microscopy.**

Topological states of matter harbor strikingly unique boundary states, such as the chiral edges of the quantum Hall effect [11], the surface states of topological insulators [12,13] and the Majorana end modes of topological superconductors. The properties of these surface states, e.g., gapless surface spectrum, relativistic dynamics, and evasion of localization by disorder, are determined by the topological nature of the bulk and are protected by the energy gap in the bulk's spectrum. These states cannot be realized as standalone systems, without the coupling to the topological bulk. Surprisingly, such states exist even on the two dimensional surfaces of three dimensional Weyl semi-metals [1-6], despite the absence of a bulk energy gap [14]. The defining characteristic of these states are the "Fermi arcs", which may not be realized on standalone two dimensional systems. While in two dimensional systems lines of constant energy must form closed contours in momentum space, "Fermi arcs" are open contours that emanate and end in states associated with bulk Dirac cones whose nodes are termed "Weyl points". Here we use Scanning Tunneling Microscopy (STM) and spectroscopy to study surface states in the Weyl semi-metal TaAs. Bulk and surface band structures of TaAs have been modeled [4-6] and mapped by photoemission spectroscopy [7-9] and its unique electrodynamics [15-19] has been probed in magnetotransport [20,21]. In our measurements we observe and characterize the Fermi arcs on the surface of TaAs, and distinguish them from the non-topological surface states.

High-quality single-crystals of TaAs (see methods section) were cold-cleaved at 80K in ultra-high vacuum exposing a fresh (001) surface that was measured at 4.2K in a commercial STM

(UNISOKU). Quasiparticle Interference (QPI) patterns that elastically scattered electrons embed in the local density of states were measured in differential conductance (dI/dV) mappings. We reveal four distinct attributes of the Fermi arcs by measuring different aspects of their scattering processes: (i) their relatively isotropic QPI profile, revealed by scattering off atomic vacancies [22-24]; (ii) the linear energy dispersion from scattering off a crystallographic step-edge; (iii) their localization on the Ta layer by tracing the spatial origin of their QPI patterns with sub-atomic resolution and (iv) the weak coupling to the surface atomic structure, as opposed to the strongly coupled trivial states, deduced from the manifestation of their wavefunction structure in the QPI pattern.

(i) An atomically resolved topographic image with several vacancies is shown in Fig. 1a. The QPI patterns appear in Fig. 1b around the vacancies seen in the dI/dV map. In TaAs these QPI patterns are superimposed on a spatial density modulation commensurate with the lattice structure (inset). Fourier decomposition of the QPI pattern at a fixed energy (Fig. 1c) separates surface scattering processes according to their transferred momentum, $q$, between incoming and scattered electronic waves. We recognize three prominent QPI patterns: ellipse-shaped patterns around $G_o$ and $G_{\pm Y}$, half of bowtie-shaped patterns at $G_{\pm X}$, and portions of rounded square-shaped patterns around the four corners of the central zone (dashed lines). These QPI patterns result from inter- and intra-band scattering of non-topological surface states. To associate these QPI patterns with particular scattering processes we plot in Fig. 1d the spin selective scattering probability (SSP) calculated (methods) for the As-terminated Fermi surface of TaAs, based on its previously extracted dispersion [5,9] and spin texture [5,25] (Fig. 1e). By comparing the SSP with the central zone in the QPI map we identify the ellipse QPI with scattering within the ellipse-shaped bands (blue arrow in Fig 1e), and the square patterns (green in the SSP) with bowtie-to-ellipse scattering (green arrow). The detected splitting of the latter into two concentric squares originates from scatterings among the spin-split copies of the ellipse and bowtie bands and thus directly reflects the strong spin-orbit coupling in TaAs [26]. Absence of doubled ellipse and bowtie QPI patterns manifest the scattering protection provided by these bands' helical spin textures [25,27].

Excellent agreement between the measured QPI and SSP as well as between the measured spectrum and the calculated density of states for As-terminated surface (Fig. 1g) identifies our surface layer as As. We note, however, that intra-band scattering within the bowtie band (yellow in SSP) around $G_o$ are hardly observed in the measured QPI. Its absence, addressed below, enables first detection of the Fermi arcs around $G_o$ (magnified in Fig. 1f, left). We find two leaf-like features that peak beyond the ellipse. Quantitative agreement with the calculated SSP (Fig. 1f, right) identifies them with scattering processes between the Fermi arcs that emanate from the W2 Weyl point (defined in Fig. 1e) and the states located at their fine structured tail (Q1 in Fig. 1e). Their arc-like shape directly reflects the contour of the Fermi-arcs.

(ii) The energy dispersion of the Fermi arcs is measured by electron scattering off a crystallographic step-edge oriented 49° with respect to the crystal axis (Fig. 2a). Accordingly, the interference pattern forms approximately along the $G_o$-$G_M$ direction (Fig. 2b). The measured dI/dV line-cut (Fig. 2c) displays clear dispersing interference patterns superimposed on commensurate (inset) non-dispersing modulations. Fourier transforming this map (Fig. 2d) reveals the energy evolution of the QPI along $G_o$-$G_M$. The dispersing ellipse- and square-shaped QPI (blue and green arrows, respectively) are identified by comparison to

SSP (Fig. 2e). Between these two we observe (red arrow) a scattering signature among two Fermi arcs (Q2 in Fig. 1e). Upon increasing energy towards W2 the extent of each arc shrinks [4-9], resulting in a linear increase of the inter-arc separation corresponding to an average velocity of ~ $10^5$ m/sec per arc. At the W2 energy (2meV above Fermi energy) the inter-arc separation becomes the inter-Weyl-point separation and equals 5.4±0.1 $nm^{-1}$. Both values are consistent with our band structure modeling [5] and photoemission spectroscopy [7-9].

(iii) In contrast to trivial states that are bound to the surface by the local surface potential, the Fermi arcs existence is guaranteed by the bulk topology. We examined the distribution of the two types of bands with respect to the topmost As layer. To that end, we decompose (methods) the dI/dV map to sub-maps measured on As sites (Fig. 2f) and on Ta sites, a monolayer deeper (Fig. 2g). Remarkably, the Fourier transforms of the two sub-maps display distinct patterns. QPI on the As-layer matches that of the ellipse band (blue arrows in Fig. 2e,h). In contrast, on the Ta-layer we find two opposite V-shaped curves (red arrows in Fig. 2i) that agree with SSP of intra-Fermi-arc processes (Q3 in Fig. 2j). Intriguingly, due to the shrinking extent of the Fermi arcs in momentum space towards the Weyl point the upper dispersing branch extrapolates to the energy of the W2 Weyl point. The distinct QPI patterns assert that the Fermi-arc's wavefunction is profoundly different from the non-topological dangling bond bands. While the latter are confined to the As termination layer, the Fermi arcs states, which relate to the bulk Weyl cone Ta states (Fig. 2h,i insets), indeed reside on the Ta sites and extend further into the bulk.

(iv) We now show that the Fermi-arc bands differ from the non-topological ones also in their structure parallel to the surface within the unit cell. Figure 3 shows QPI maps at three different energies (right panels) alongside dI/dV maps in a vacancy-free region (left panels). At -300 meV (below Fermi energy) the vacancy-free dI/dV map and the corresponding QPI map are approximately symmetric to $90^0$ rotations and the QPI patterns are concentrated around $G_o$. In contrast, at 85 meV, 130 meV and at the Fermi energy (Fig. 1b inset), the vacancy-free dI/dV shows clear chains that change their crystallographic orientation with energy, while the QPI features are strongly replicated along that modulation direction. Modulation in a *vacancy-free* region ought to be attributed to the structure of the wavefunction. The correlation demonstrated in Fig. 3 is established in Fig. 4a by comparing the intensities of the different features. The intensity of vacancy-free dI/dV modulations along Γ-Y is fully correlated with the $G_o$–$G_Y$ replications of the ellipse's QPI, and similarly for the bowtie along the Γ-X direction. The strong correlation between the two seemingly unrelated phenomena extends to all energies and suggests that both are dictated by the structure of the wavefunction rather than the details of the scatterer. In contrast to the trivial bands the Fermi arcs' QPI does not show any detectable replications implying their relatively uniform distribution within the unit cell

Aiming at using this distinction to separate the different states, we note that both the dI/dV map at the vacancy-free regions and the QPI near vacancies reflect the coupling of the electrons to the periodic potential in the surface plane. Bloch theorem constrains a state with a crystal momentum $k$ to be a super-position of momenta $k + G$, where $G$ is a vector in the two dimensional reciprocal lattice, $\Psi_k(r) = \sum_G C_G^k e^{i(k+G)\cdot r}$. Consequently, the local density of states in a vacancy-free region becomes $\sum_g A_g e^{ig\cdot r}$ where $A_g = \sum_{G,k} C_G^{*k} C_{G'}^k \delta(E - E_k)$ is the amplitude of the Bragg peak that corresponds to $g = G - G'$, r is the position, E is the energy and $E_k$ the energy of the state with momentum k. A state with multiple significant Bloch coefficients $C_G^k$ has a fine structure within the unit cell, which translates to multiple Bragg peaks. A vacancy violates the periodicity and adds a potential *V(r)*, whose Fourier transform is $V_q$. The vacancy may scatter an electron between states $\Psi_k(r)$ and $\Psi_{k'}(r)$ through any momentum transfer $q_g$ satisfying $q_g = k - k' + g$. The amplitude for each of these processes is proportional, within the Born approximation, to $\sum_{G'} V_{q_g} C^{*k}_{G'+g} C^{k'}_{G'}$. Hence, multiple significant coefficients $C^k_G$ result in replicas of the QPI around multiple Bragg peaks, limited by the ability of the potential to provide the required momentum transfer.

Accordingly, the replicated QPI observed in Fig. 3 originates from bands whose wavefunctions include several significant Bloch components. We attempt to eliminate these states from the QPI map by subtracting off their scaled replicated signals from the $G_o$ signal. In Fig. 4b we show the outcome of subtracting the ellipse QPI around $\pm G_Y$ from that around $G_o$ at $E_F$. While the ellipse is eliminated, the Γ-Y Fermi-arcs' QPI signature remains unchanged. Remarkably, this elimination further exposes a signature of the Γ-X Fermi arcs (Compare to SSP of Fermi-arcs shown in Fig. 4c). This observation indicates that the Fermi-arc wave function on the surface is composed of a single dominant term ($g = 0$) or of combination of terms whose momentum difference ($g = G - G'$) is larger than our resolution.

We expect this analysis technique of the replicated structure of QPI patterns for separating overlapping features to be of further applicability in future studies of Fermi arcs in Weyl semi-metals. On a broader scope, numerous electronic systems have been studied using QPI measurements. Some of these exhibit clear QPI replications. Two prime examples are topological crystalline insulators [28], in which the topological surface states are replicated, and high temperature superconductors [29,30], in which charge order has been recently reported. It would be enticing to apply this analysis to characterize the structure of the Bloch wave functions in these systems and to unveil hidden spectroscopic features.

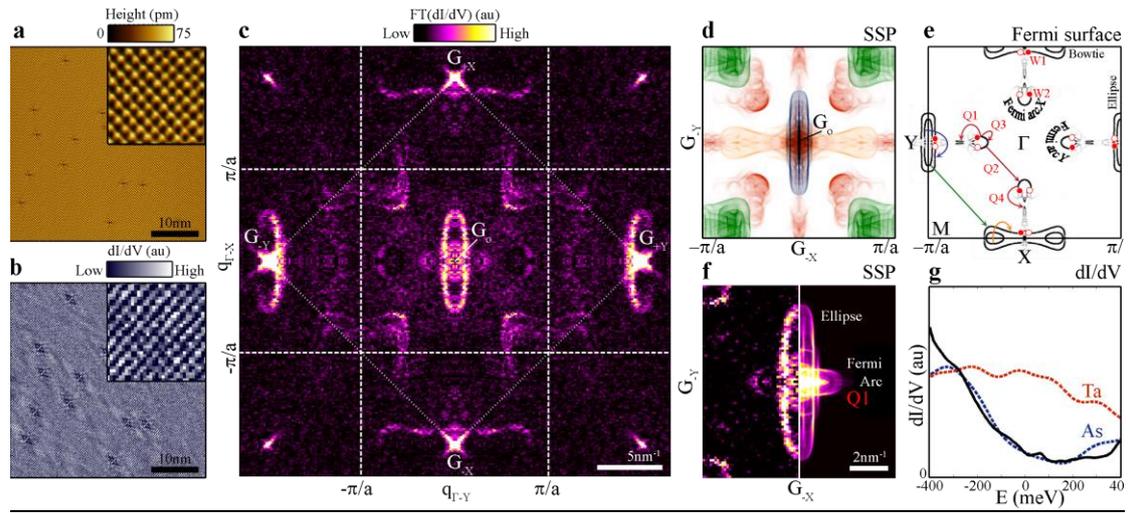

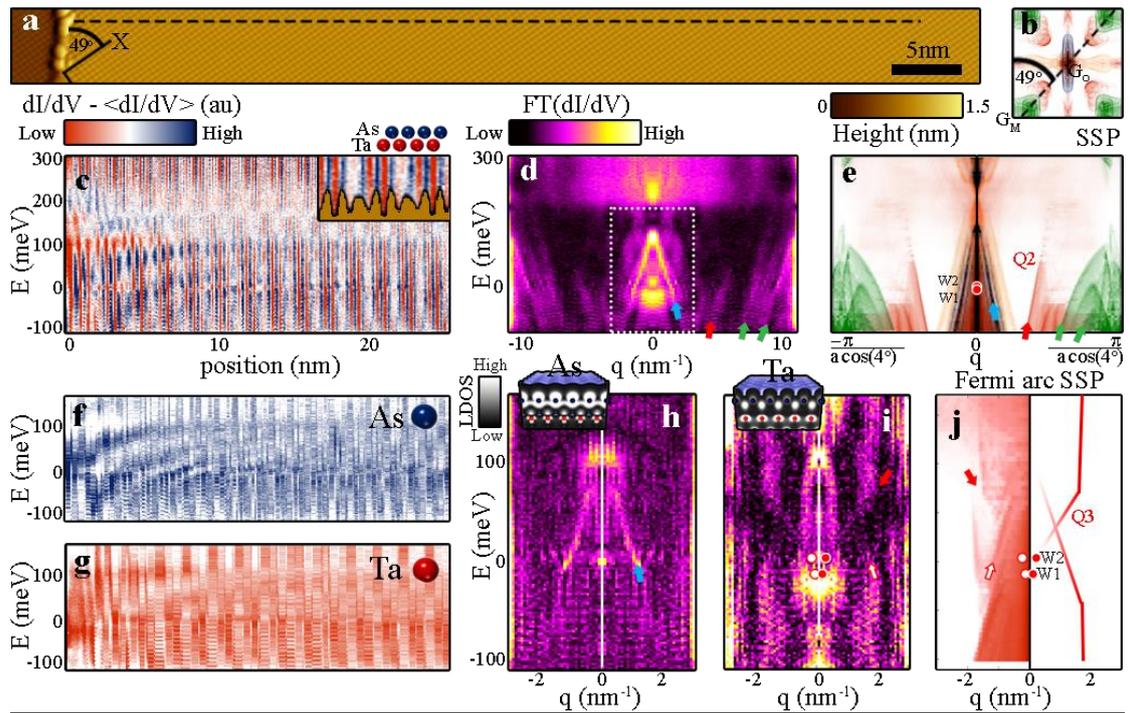

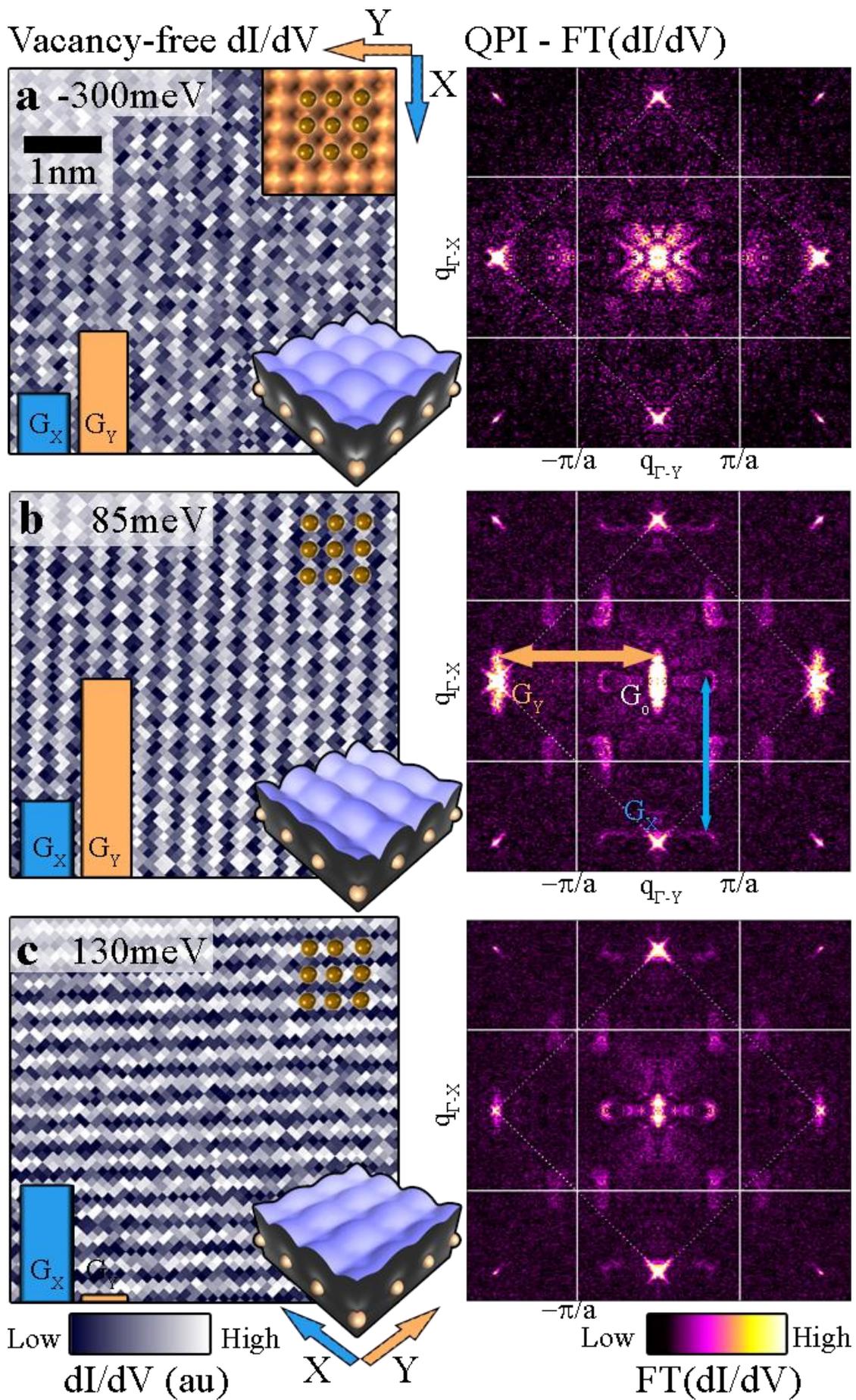

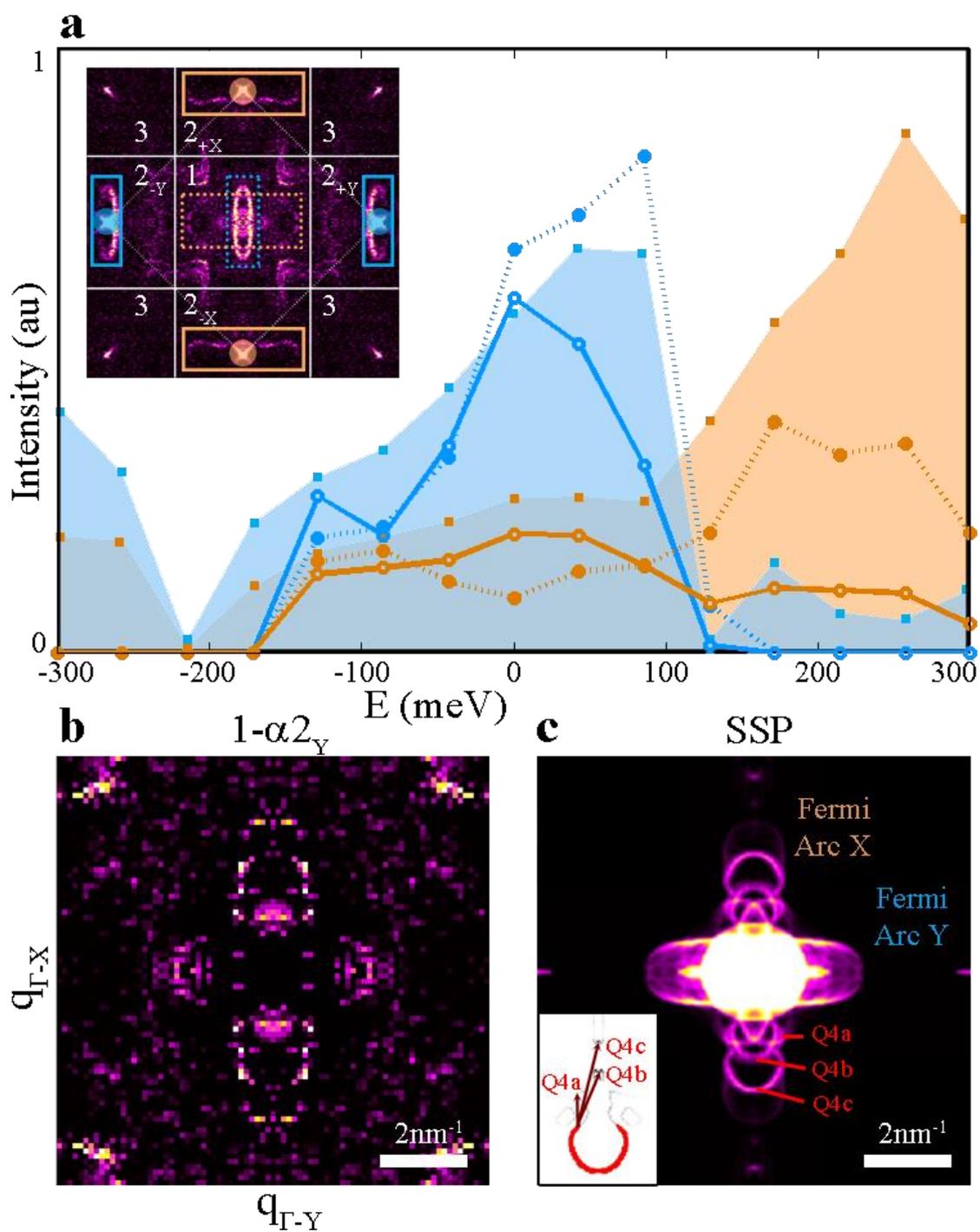

## Figure captions

**Figure 1. Quasi-particle interference of trivial bands and Fermi arcs in TaAs. a** Topography ($V_B$=-400meV, $I_t$=350pA) of the (001) surface of TaAs with a few As-vacancies. Inset: atomic resolution with a lattice constant a=3.47Å. **b** dI/dV map of a ($V_B$=0meV) shows elongated QPI about vacancies on top of an atomically modulated local density of states (inset). **c** Fourier analysis of b reveals an intricate QPI pattern centered around both q=0 and Bragg peaks. **d** DFT calculated Fermi surface of As terminated TaAs contains bowtie and ellipse shaped bands and topological Fermi arcs emanating from Weyl points W1,W2 at energy -13 meV, 1.5 meV respectively (red). Intra- and inter-band scattering processes are represented by colored arrows. Intra- (Q1,Q3,Q4) and inter-Fermi arc (Q2) processes are labeled. **e** SSP based on d shows remarkable resemblance to central zone in c: ellipse (blue) and bowtie (yellow) shaped patterns, spin-orbit-split double squares (green), intra-Fermi arc scattering around $G_o$ (i.e., *q=0*) and inter-Fermi arc along $G_o$-$G_M$. Some SSP features appear in c both around $G_o$ and around Bragg peaks $\pm G_X$, $\pm G_Y$. **f** A magnified view of the $G_o$ vicinity (left) reveals a round arc beyond the ellipse pattern (left). It agrees well with the SSP of scattering about the Fermi arc (Q1) and that of the ellipse (right). **g** Measured dI/dV spectrum (solid) and calculated density of states of As and Ta terminated (001) surfaces (blue and red, respectively), indicating the As termination of the measured sample.

**Figure 2. Fermi arcs on Ta, Trivial bands on As. a** Step-edge Topography oriented 49° with respect to the crystal axis. **b** Corresponding 49° cut (dashed line) across the SSP. **c** dI/dV map (along the dashed line in a) shows modulations in the density of states. Inset shows its commensuration with the topographic profile. It exhibits high density of states on As atoms and lower on Ta, as well as dispersive interference patterns. **d** Fourier analysis of c shows the energy dispersion of the ellipse (blue arrow), the squares (green arrows) and the Fermi arcs (Q2 in Fig. 1e). **e** SSP cut along the dashed line in b identifies Q2 with inter-Fermi arc scattering, extending above and below the Weyl points' energies ($E_{W2}$~2meV, $E_{W1}$~ -13meV, respectively).**f,g** density of states on As sites (blue) and on Ta sites (red) reveals two different dispersing modes. **h** Fourier analysis of f detects the energy dispersion of the ellipse QPI. Inset: calculated electronic density of states of the ellipse band shows high localization on As sites. **i** Fourier analysis of g unveils QPI pattern, associated with the Fermi arcs, which extrapolates to W2. Inset: Fermi arc wavefunction is localized on Ta sites. **j** left: SSP of Fermi arcs only, right: dispersion extracted from left panel. Note the resemblance to h. One of the modes extrapolates to the projection of W2 due to shrinkage of the Fermi arc (d, inset)).

**Figure 3. Correlation between Modulations and Replications set by the electronic wavefunction. a-c, Left panels:** dI/dV in a vacancy-free region at three representative energies shows strong modulation whose strength and orientation change with energy. The bars represent the Bragg peak intensities, $\left|A_{G_X,G_Y}\right|^2$, along the two crystallographic directions. Insets: DFT calculation of the Local charge density captures a similar modulation. **a-c, right panels:** Fourier analysis of extended dI/dV maps in the presence of vacancies at the corresponding energies. The intensity and anisotropy of the replications of QPI features are correlated with the modulation detected in the vacancy-free region.

**Figure 4. Replicated QPI patterns as a spectroscopic tool. a** Energy dependence of the Bragg peaks' intensities at the vacancy-free region (shaded) along with the intensities of the bowtie and ellipse QPI patterns (orange and blue, respectively) around q=0 and around the

Bragg peaks (see inset for legend). The ellipse is correlated with the Bragg peaks along $\Gamma$-Y, while the bowtie is correlated with Bragg peaks along $\Gamma$-X. **b** Subtraction of zones $2_{+Y}$ and $2_{-Y}$ times $\alpha=1.14$ (extracted from a, details in supplementary) from the central zone eliminates the ellipse from around q=0 while leaving the QPI pattern of the $\Gamma$-X Fermi arc unaffected. Signatures of scatterings among the $\Gamma$-Y Fermi arc are revealed. **c** SSP of Fermi arcs alone. Inset: Contributing scattering processes (Q4a-c) within the Fermi arc along $\Gamma$-Y.

References:

1. Wan, X., Turner, A. M., Vishwanath, A. & Savrasov, S. Y. Topological semimetal and Fermi-arc surface states in the electronic structure of pyrochlore iridates. *Phys. Rev. B* **83,** 205101 (2011).
2. Xu, G., Weng, H., Wang, Z., Dai, X. & Fang, Z. Chern semimetal and the quantized anomalous Hall effect in $HgCr_2Se_4$. *Phys. Rev. Lett.* **107,** 186806 (2011).
3. Burkov, A. A. & Balents, L. Weyl semimetal in a topological insulator multilayer. *Phys.Rev.Lett.* **107**, 127205 (2011).
4. Weng, H., Fang, C., Fang, Z., Bernevig, B. A. & Dai, X. Weyl semimetal phase in noncentrosymmetric transition-metal monophosphides. *Phys. Rev. X* **5**, 011029 (2015).
5. Sun, Y., Wu, S.C., Yan, B. Topological surface states and Fermi arcs of the noncentrosymmetric Weyl semimetals TaAs, TaP, NbAs, and NbP. *Phys. Rev. B* **92**, 115428 (2015).
6. Huang, S.M. *et al.* An inversion breaking Weyl semimetal state in the TaAs material class, Nat. Commun. **6**, 7373 – 7379 (2015).
7. Xu, S.Y. *et al.* Discovery of a Weyl fermion semimetal and topological Fermi arcs. *Science* **349,** 613–617 (2015).
8. Lv, B. Q. *et al.* Experimental discovery of weyl semimetal TaAs. *Phys. Rev. X* **5,** 031013 (2015).
9. Yang, L. X. *et al.* Weyl semimetal phase in the non-centrosymmetric compound TaAs. *Nat. Phys.* **11,** 728–732 (2015).
10. Zheng H. *et al.* Atomic-Scale Visualization of Quantum Interference on a Weyl Semimetal Surface by Scanning Tunneling Microscopy. *ACS Nano* **10**, 1378 (2016)
11. Wen, X. G. Int. J. Mod. Phys. B 4, 239–271 (1990).
12. Hasan and M. Z., Kane, C. L. Kane, Colloquium: Topological Insulators, *Rev. Mod. Phys.* **82**, 3045 (2010).
13. Qi, X. L. and Zhang, S.-C., Topological Insulators and Superconductors, *Rev. Mod. Phys.* **83**, 1057 (2011).
14. Murakami, S. Phase transition between the quantum spin Hall and insulator phases in 3D: Emergence of a topological gapless phase. *New J. Phys.* **10**, 029802 (2008).
15. Vazifeh, M. M. & Franz, M. Electromagnetic Response of Weyl Semimetals. *Phys. Rev. Lett.* **111**, 027201 (2013)
16. D. T. Son and B. Z. Spivak. Chiral anomaly and classical negative magnetoresistance of Weyl metals. *Phys. Rev. B* **88**, 104412 (2013)
17. Potter, A. C., Kimchi,I., & Vishwanath, A. Quantum oscillations from surface Fermi arcs in Weyl and Dirac semimetals. *Nat. Commun.* **5**, 5161 (2014)
18. Baum, Y., Berg, E., Parameswaran, S. A. & Stern, A. Current at a Distance and Resonant Transparency in Weyl Semimetals. *Phys. Rev. X* **5**, 041046 (2015).
19. Baireuther, P., Hutasoit, J. A., Tworzydło, J., Beenakker, C. W. J. Scattering theory of the chiral magnetic effect in a Weyl semimetal: Interplay of bulk Weyl cones and surface Fermi arcs. arXiv: 1512.02144.
20. Zhang, C. *et al.* Observation of the Adler-Bell-Jackiw chiral anomaly in a Weyl semimetal. arXiv 1503.02630.
21. Huang, X. *et al.* Observation of the Chiral-Anomaly-Induced Negative Magnetoresistance in 3D Weyl Semimetal TaAs. *Phys. Rev. X* **5,** 031023 (2015).
22. Chang, G. *et al.* Signatures of Fermi Arcs in the Quasiparticle Interferences of the Weyl Semimetals TaAs and NbP. *Phys. Rev. Lett.* **116,** 066601 (2016).
23. Belopolski, I. *et al.* Criteria for Directly Detecting Topological Fermi Arcs in Weyl Semimetals. *Phys. Rev. Lett.* **116,** 066802 (2016).

24. Li, J., Wang, Z., Yazdani, A., & Bernevig B.A., Universal signatures of Fermi arcs in quasiparticle interference on the surface of Weyl semimetals. *Phys. Rev. B,* **93**, 041109 (2016)
25. Lv B.Q. *et al.* Observation of Fermi-Arc Spin Texture in TaAs *Phys. Rev. Lett.* **115** 217601 (2015)
26. Z. K. Liu, *et al.* Evolution of the Fermi surface of Weyl semimetals in the transition metal pnictide family. *Nature Materials*, **15**, 27–31 (2016)
27. Roushan, P. *et al.* Topological surface states protected from backscattering by chiral spin texture. *Nature,* **460**, 1106 (2009)
28. Zeljkovic , I, *et al.*, Mapping the unconventional orbital texture in topological crystalline insulators *Nat. Phys.*, **10**, 572–577 (2014)
29. J. E. Hoffman J.E., *et al.* Imaging Quasiparticle Interference in Bi2Sr2CaCu2O8+x, *Science* **297**, 1148-1151 (2002)
30. Dalla Torre, E.G., He, Y & Demler, E. Holographic Maps of Quasiparticle Interference. arXiv: 1512.03456

**Acknowledgements:**

H.B. acknowledges support from the European Research Council (Starter Grant No. 678702 "TOPO-NW"), the ISF, and the U.S.-Israel BSF
C.F. acknowledges support from the European Research Council (Advanced Grant No. 291472 "Idea Heusler").
A.S. acknowledges support from the European Research Council under the European Union's Seventh Framework Programme
(FP7/2007-2013) / ERC Project MUNATOP, Minerva foundation, and the U.S.-Israel BSF